\documentclass[pdflatex]{article}
\usepackage{spconf,amsmath,graphicx}

\title{SVGEDITBENCH V2: A BENCHMARK FOR INSTRUCTION-BASED SVG EDITING}
\name{Kunato Nishina, Yusuke Matsui}
\address{The University of Tokyo, Japan}

\usepackage{url}
\usepackage{bm}
\usepackage{autobreak}
\usepackage{multirow}
\usepackage{amssymb}
\usepackage{booktabs}

\newcommand{\eg}{\emph{e.g.}}

\newcommand{\etal}{\emph{et al.}}
\newcommand{\etc}{\emph{etc.}}

\newcommand{\Fref}[1]{Fig. \ref{#1}}
\newcommand{\Sref}[1]{Sec. \ref{#1}}
\newcommand{\Tref}[1]{Tab. \ref{#1}}

\usepackage[dvipsnames]{xcolor}

\usepackage{subfig}
\usepackage{listings}
\usepackage{ascmac}
\begin{document}
\maketitle
\begin{abstract}
Vector format has been popular for representing icons and sketches. It has also been famous for design purposes. Regarding image editing, research on vector graphics editing rarely exists in contrast with the raster counterpart. We considered the reason to be the lack of datasets and benchmarks. Thus, we propose SVGEditBench V2, a benchmark dataset for instruction-based SVG editing. SVGEditBench V2 comprises triplets of an original image, a ground truth image, and the editing prompt. We built the dataset by first extracting image pairs from various SVG emoji datasets. Then, we had GPT-4o to create the prompt. We found that triplets gained by this simple pipeline contain varying sorts of editing tasks. Additionally, we performed the editing tasks with existing LLMs and investigated how those current methods can perform SVG editing. Although there were some successful cases, we found that there is a massive room for improvement.
\end{abstract}
\begin{keywords}
  SVG, vector graphics, image editing, emojis, Large Language Models
\end{keywords}

\section{INTRODUCTION}
\label{sec:introduction}

\begin{figure*}
  \centering
  \includegraphics[width=0.9\hsize]{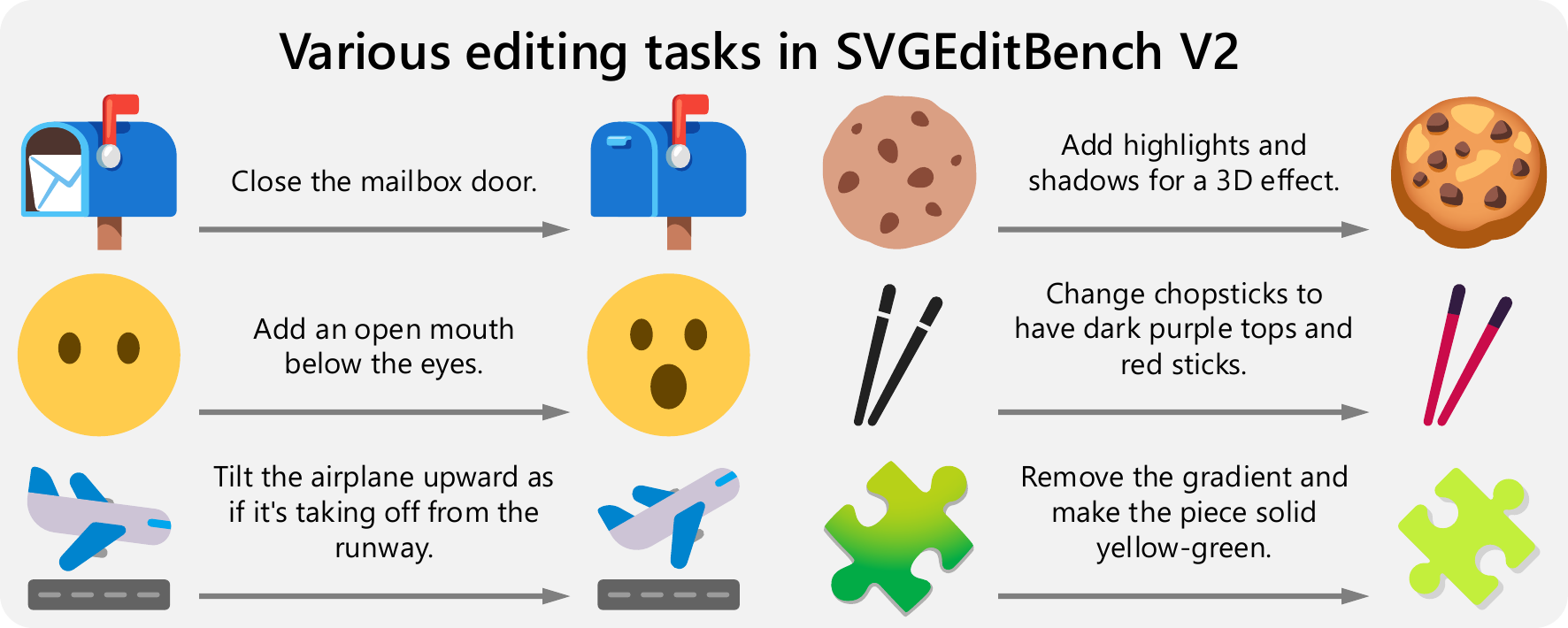}
  \caption{SVGEditBench V2 is a collection of triplets consisting of SVG graphics before and after editing and the editing prompt. We extracted the images from multiple publicly available emoji datasets and employed GPT-4o to generate the prompts. This straightforward pipeline can produce a variety of editing tasks. The tasks range from transforming elements to changing the overall style. A high-level understanding of the graphics is necessary to perform these tasks.}
  \label{fig:teaser}
\end{figure*}
With the advancements of Diffusion models, recent research~\cite{stable-diffusion,InstructPix2Pix} attempts to generate or edit raster images according to an input text. Also, generating vector graphics---an alternative way to represent images---has become increasingly popular~\cite{vectorfusion}.

Despite this, not much research exists in text-to-vector editing. Zhang \etal ~\cite{svg-customization} proposes a method to customize an SVG image with a text prompt. However, the method requires fine-tuning a Diffusion model for each exemplar image, which is costly. Also, some research~\cite{SVG-LLM, vgbench, SVGEditBench} aims to generate or understand vector images with Large Language Models (LLMs). They focus on the fact that vector graphics are human-readable text data. However, none of these proposed a new vector graphics editing method.

We believe the limited research on vector graphics editing is due to the absence of unified benchmarks and datasets. Thus, we propose SVGEditBench V2, a benchmark dataset for evaluating methods for editing SVGs. SVGEditBench V2 comprises 1,683 triplets (examples in \Fref{fig:teaser}). Each triplet includes the original SVG image before editing, the ground truth SVG image after editing, and the editing text prompt. The editing prompt describes the changes from the original image to the ground truth. We extracted the images from public SVG emoji datasets and asked GPT-4o to generate the prompt using those images. When evaluating SVG editing methods, we first edit the original image using the editing prompt. We then compare the output image from the method with the ground truth using raster-based, contour-based, and description-based metrics. We will make the created dataset publicly available on GitHub.

We applied our benchmark to existing 15 LLMs and 7 Large Multimodal Models (LMMs) to examine their ability for SVG editing. Our experiments revealed that, generally, these models struggle to edit images following the prompts. Still, there were a few instances where the models succeeded in their editing tasks when they were simple.
\section{RELATED WORK}
This section introduces prior work that inspired our dataset creation and model evaluation pipelines. We focus on two aspects: benchmarks in vector graphics processing and datasets on image editing.

\subsection{Benchmarks on vector graphics processing}
Several benchmarks have been proposed for vector graphics processing. SVGBench~\cite{starvector} proposes an evaluation method for image vectorization by extracting images from existing datasets and defining metrics. VGBench~\cite{vgbench} is a benchmark for quantitatively analyzing the performance of VQA and generation with three vector graphics formats: SVG, TikZ, and GraphViz. The existing research closest to ours is SVGEditBench~\cite{SVGEditBench}. SVGEditBench is a dataset built to evaluate the SVG editing capabilities of LLMs. 

Our benchmark is different from these previous works in the following points. 
\begin{itemize}
  \item SVGEditBench V2 assesses the performance of SVG \textbf{editing}. To our knowledge, no other benchmark exists besides SVGEditBench.
  \item SVGEditBench V2 offers a wider range of complex editing tasks than SVGEditBench. SVGEditBench is restricted to six predefined and easy tasks. Also, the benchmark is not applicable for other models than LLMs since straightforward heuristics can address these tasks. However, our dataset is not confined to a specific set of tasks. These tasks require a semantic understanding of the input image. For example, a task may require identifying which part of the SVG code corresponds to the object specified.
  \item SVGEditBench V2 evaluates SVG editing from a broader perspective. SVGEditBench primarily uses Mean Squared Error (MSE) as the evaluation metric. However, semantic and geometric aspects are also important when assessing SVG editing. Therefore, we utilize four distinct metrics to encompass these areas during the evaluation phase.
\end{itemize}

\subsection{Similar works in the raster domain}
Datasets for image editing for arbitrary tasks also exist in the raster domain. The representative dataset is the one InstructPix2Pix (IP2P)~\cite{InstructPix2Pix} used to train. In IP2P, a fine-tuned GPT-3 generated an appropriate editing instruction and the image caption after the edit. Then, they employ Prompt-to-Prompt~\cite{prompt-to-prompt} to generate a corresponding image consistent with the original, and these images collectively constitute a dataset suitable for training an image editing model.

Subsequent research has proposed image editing techniques and datasets~\cite{magicbrush, SmartEdit} based on IP2P. They pointed out that the images may not correspond to the instruction since IP2P uses automatically generated images~\cite{magicbrush} and that IP2P cannot handle complex instructions requiring reasoning~\cite{SmartEdit}. The papers address these problems with manual annotation or filtering~\cite{magicbrush, SmartEdit}, or with segmentation of the objects in the image~\cite{SmartEdit}. In our dataset, we generated the editing prompt with an LLM inspired by IP2P.

\section{SVGEDITBENCH V2}
\label{sec:method}
\begin{figure*}
  \centering
  \includegraphics[width=0.9\hsize]{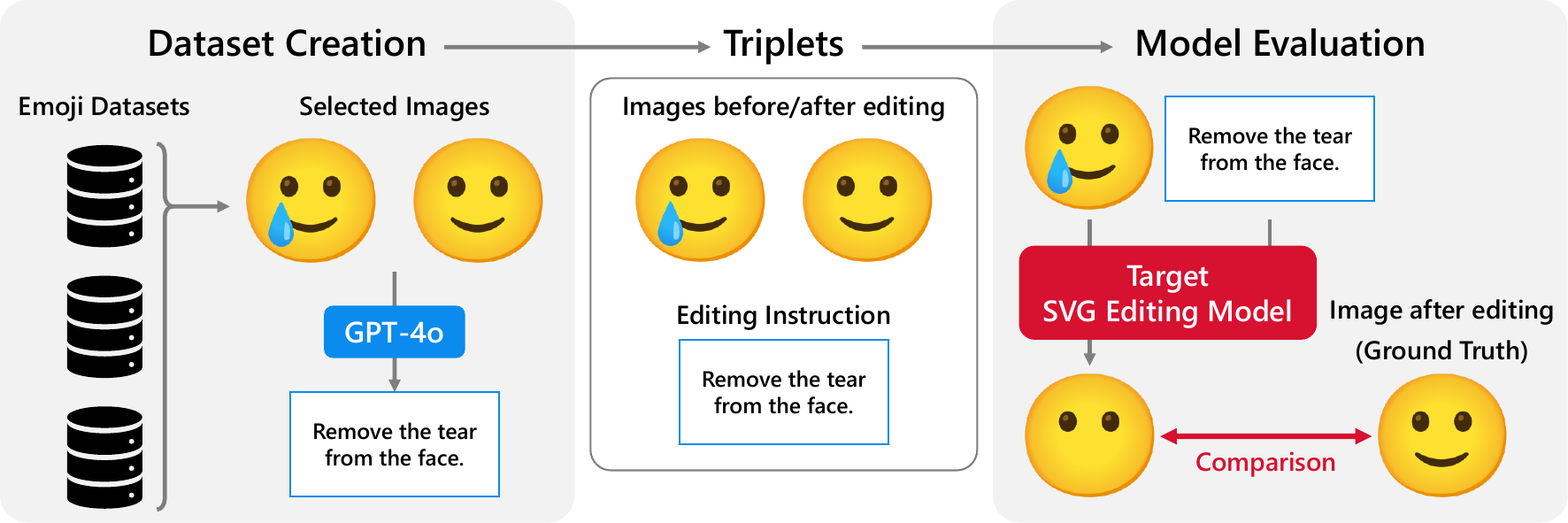}
  \caption{Overview of our method. Our benchmarking pipeline comprises two main parts: dataset creation and model evaluation. Refer to \Sref{sec:method} and the supplementary material for more details.}
  \label{fig:overview}
\end{figure*}

This section briefly describes each component in the dataset creation and evaluation pipelines in SVGEditBench V2. \Fref{fig:overview} shows the overview of our proposed approach. We provide further details in the supplementary material. Note that although we used emojis as the images, similar pipelines are applicable when creating editing benchmarks for other vector graphics domains.

\subsection{Selection of original images}
Inspired by SVGBench~\cite{starvector} and SVGEditBench~\cite{SVGEditBench}, we built the dataset from existing SVG emoji datasets: Noto Emoji~\cite{noto-emoji}, Twemoji~\cite{twemoji}, and flat and high-contrast versions of Fluent Emoji~\cite{fluent-emoji}. The reasons for choosing emojis are as follows.
\begin{itemize}
    \item Emoji datasets are suited for generating editing tasks. We can obtain images of similar layouts but in different styles when we pick images with the same Unicode codepoint from different datasets. This fact leads to the inclusion of style-transfer tasks. Also, emojis contain multiple images of the same type, such as facial expressions and clocks. With this feature, the editing tasks can better align with real-life editing tasks.
    \item We can use emoji names defined in Unicode to describe the images. These descriptions can help extract suitable image pairs and generate accurate editing instructions in subsequent steps.
\end{itemize}

We filtered out some images from those datasets. Specifically, we excluded images that may be inappropriate for editing and images we could not obtain the description. The number of remaining images was 5,766.

\subsection{Extraction of image pairs}
We extracted an original image (before editing) and a ground truth image (after editing) from these selected image sets. The image pair should look similar and have close meanings. This characteristic makes the tasks realistic and allows the editing prompt to describe the differences clearly. Also, the dataset as a whole should contain diverse images and editing tasks.

We calculated a distance metric for all combinations of two images in the image set and picked image pairs with the closest distance. However, to prevent the same image from being used too frequently, we allowed the same image to be in the dataset only up to three times. The metric here is a sum of LPIPS~\cite{lpips} distance and the similarity of the descriptions measured by the CLIP~\cite{clip} text embeddings. This metric is suitable for obtaining image pairs with the attributes mentioned earlier because it considers both the visual and semantic aspects of the images. We extracted 3,000 image pairs with this process.

\subsection{Generation of editing instruction}
The next step is to create the editing instructions for each image pair. Inspired by InstructPix2Pix~\cite{InstructPix2Pix}, we generated the instructions using GPT-4o (\texttt{gpt-4o-2024-08-06}). We provided GPT-4o the rasterized versions of the original and the ground truth images. We also included the descriptions for each image in the prompt. This strategy differs from IP2P since they made the instructions only from text input.

\subsection{Filtering the triplets}
Up to this point, we gained triplets of an original image, a ground truth image, and an editing instruction that links them. However, these instructions may be inaccurate. Hence, we manually checked them and removed inappropriate ones to reduce noise. We filtered the triplets from two aspects. Firstly, the prompt should mention all the elements in the image that should be changed. Secondly, the prompt should not contain errors that would significantly impact the editing results.

Additionally, we filtered out the triplets if the prompt in the evaluation phase ($\approx$ instruction + original SVG code) would exceed 5,000 characters. Since the LLMs used in the following experiment may not fit long prompts within their context, this process was necessary to reduce the impact of differences in context length on the evaluation results. As we used existing SVG data, we believe this process should not significantly affect the insights on LLMs' ability to edit SVG images. The number of the remaining triplets became 1,683.

\subsection{Analysis of the created dataset}
\Fref{fig:teaser} lists some triplets created by the above pipeline. Compared to SVGEditBench, our tasks are much more complex and diverse. Firstly, the editing prompt specifies the objects the editing model should modify by their names, not color, as in SVGEditBench. It is not straightforward to find the specified object from the SVG code. Also, the difficulty of the tasks varies remarkably from one another. Some are simple change-color tasks, as in the middle-right example of \Fref{fig:teaser}. In contrast, others require generating objects that match the style of the original image and placing it appropriately (top-left example of \Fref{fig:teaser}).

\subsection{Evaluation metrics}
We now evaluate SVG editing techniques (including LLMs) with our created dataset. Given a target SVG editing method, we edit the original image from each triplet using the associated editing prompt. We compare the output with the ground truth image in the triplet. 

We used four metrics in this benchmark. We chose these metrics so that the evaluation reflects the various aspects of SVG editing. Firstly, the Mean Squared Error (MSE) of pixel values and cosine similarity of DINOv2~\cite{dinov2} features compare the images in raster format. These measure how the images are visually similar. Next, we used CLIPScore~\cite{clipscore} between the description text of the ground truth and the rasterized output image to judge the semantic closeness.

Additionally, we used the Chamfer distance metric to evaluate the geometrical closeness of the images. We used a two-step Chamfer distance to account for both the position of the shapes and the difference of the shapes themselves. In this metric, we first calculate the Chamfer distance $d_\mathrm{shape}$ between shapes in the output and ground truth images. We define the distance between the images by summing $d_\mathrm{shape}$ between a shape in one image and its closest match. More details are in the supplementary material.

We must accurately determine the contour from the SVG code to calculate this distance. Therefore, we excluded images that contain elements other than \texttt{<path>}, basic shape elements (\texttt{<rect>} \etc), and elements representing color gradients (ignored in calculating this distance).
\begin{table*}
  \small
  \renewcommand{\arraystretch}{0.7}
  \centering
  \begin{tabular}{@{}llcc|cccc|cc@{}}
      \toprule
       & & & & \multicolumn{4}{c}{Metrics} & \multicolumn{2}{c}{Error Rate} \\
       & Model & Type & \# Params & MSE \textdownarrow & DINO \textuparrow & CLIPScore \textuparrow& Chamfer \textdownarrow & Extraction \textdownarrow & Parsing \textdownarrow \\ \midrule
       & Original & & & $0.061$ & $0.897$ & $26.882$ & $2.584\times 10^3$ & & \\
       & Ground Truth & & & $0$ & $1$ & $27.492$ & $0$ & & \\ \midrule
      \multirow[t]{15}{*}{Open}
       & CodeGemma& Code & 8.54B & $0.084$ & $0.828$ & $26.281$ & $2.596\times 10^3$ & $9.7\%$ & $2.6\%$\\
       & Code Llama& Code & 6.74B & $0.103$ & $0.775$ & $25.714$ & $3.941\times 10^3$ & $14.1\%$ & $5.3\%$\\
       & Gemma 1.1 & & 8.54B & $0.077$ & $0.853$ & $26.542$ & $3.109\times 10^3$ & $1.6\%$ & $9.7\%$\\
       & Gemma 2& & 2.61B & $0.087$ & $0.837$ & $26.072$ & $2.521\times 10^3$ & $33.6\%$ & $7.4\%$\\
       & Gemma 2 & & 9.24B & $0.076$ & $0.866$ & $26.369$ & $\mathbf{2.411\times 10^3}$ & $37.3\%$ & $1.8\%$\\
       & Llama 2& & 6.74B & $0.102$ & $0.791$ & $25.686$ & $2.946\times 10^3$ & $33.1\%$ & $14.7\%$\\ 
       & Llama 3& & 8.03B & $0.099$ & $0.808$ & $26.144$ & $3.548\times 10^3$ & $6.5\%$ & $13.0\%$\\ 
       & Llama 3.1& & 8.03B & $0.108$ & $0.792$ & $26.015$ & $5.413\times 10^{27}$ & $7.1\%$ & $14.5\%$\\ 
       & Llama 3.2& & 3.21B & $0.119$ & $0.764$ & $25.762$ & $2.999\times 10^3$ & $11.6\%$ & $34.8\%$\\ 
       & LlaVa-NeXT& MM & 7.57B & $0.110$ & $0.742$ & $25.401$ & $1.476\times 10^4$ & $20.7\%$ & $22.3\%$\\ 
       & Mistral v0.2& & 7.24B & $0.115$ & $0.736$ & $25.474$ & $5.500\times 10^3$ & $9.4\%$ & $18.8\%$\\
       & Mistral v0.3& & 7.25B & $0.109$ & $0.771$ & $25.760$ & $4.311\times 10^3$ & $4.1\%$ & $10.2\%$\\
       & Phi-3.5-mini& & 3.82B & $0.128$ & $0.701$ & $25.295$ & $7.873\times 10^4$ & $66.2\%$ & $25.3\%$\\ 
       & Phi-3.5-vision& MM & 4.15B & $0.165$ & $0.570$ & $24.326$ & $3.446\times 10^3$ & $89.9\%$ & $61.8\%$\\ 
       & Qwen2-VL& MM & 8.29B & $0.104$ & $0.770$ & $25.592$ & $2.935\times 10^3$ & $21.7\%$ & $21.1\%$\\ \midrule
       \multirow[t]{7}{*}{Closed}
       & Gemini 1.0 Pro& & & $0.076$ & $0.857$ & $26.287$ & $2.876\times 10^3$ & $39.5\%$ & $\mathbf{0.6\%}$\\
       & Gemini 1.5 Flash& MM & & $\mathbf{0.070}$ & $0.869$ & $26.724$ & $2.566\times 10^3$ & $3.4\%$ & $0.8\%$\\ 
       & Gemini 1.5 Pro& MM & & $0.077$ & $0.854$ & $26.480$ & $3.370\times 10^3$ & $0.1\%$ & $0.8\%$\\ 
       & GPT-3.5 & & & $0.073$ & $0.867$ & $26.678$ & $3.733\times 10^3$ & $1.5\%$ & $1.1\%$\\ 
       & GPT-4o & MM & & $\mathbf{0.070}$ & $0.874$ & $26.784$ & $2.930\times 10^3$ & $0.1\%$ & $\mathbf{0.6\%}$\\
       & GPT-4o mini & MM & & $\mathbf{0.070}$ & $\mathbf{0.875}$ & $26.761$ & $3.482\times 10^3$ & $\mathbf{0.0\%}$ & $2.1\%$\\ 
       & o1-mini & & & $0.073$ & $0.872$ & $\mathbf{26.788}$ & $4.943\times 10^3$ & $0.3\%$ & $0.9\%$\\
      \bottomrule
  \end{tabular} %
  \caption{Results of evaluating current LLMs/LMMs' SVG editing ability with our proposed benchmark dataset. ``MM'' refers to multimodal models, and ``Code'' refers to code-specific models. The ``\# Params'' column for the open models shows the number of parameters stated on the corresponding HuggingFace page.}
  \label{tab:main-results}
\end{table*}

\section{EXPERIMENT}
In this section, we utilized current LLMs and LMMs as SVG editors and assessed their performance using SVGEditBench V2. These models can serve as a valuable baseline for SVG editing. While editing vector graphics requires professional skills, the communication ability of these models can simplify the process by offering a text-based interface, thus lowering the barrier to entry.

\subsection{Experimental setting}
We performed evaluations using various LLMs and LMMs, as shown in \Tref{tab:main-results}. The purpose of using LMMs is to confirm whether their vision capabilities can help with SVG editing. To measure the pure capabilities of LLMs and LMMs, we used simple prompts for inference, and we did not perform any prompt engineering, such as showing examples. Even for LMMs, we did not input images. Please refer to the supplementary material for the actual prompt. For open-sourced models, we used the versions with instruction-following capabilities with under 10B parameters. When running these models, we used the vLLM library~\cite{vllm} for efficient inference.

The LlaVa-NeXT model here leverages Mistral v0.2. The GPT-3.5, GPT-4o, GPT-4o mini, and o1-mini models are the ones available as \texttt{gpt-3.5-turbo-0125}, \texttt{gpt-4o-2024-08-06}, \texttt{gpt-4o-mini-2024-07-18} and \texttt{o1-mini-2024-09-12} in the OpenAI API.

\subsection{Results}
\Tref{tab:main-results} shows the results of evaluating the LLMs and LMMs with SVGEditBench V2. These are the metrics averages calculated for all the triplets whose metrics were successfully calculated. On the top of \Tref{tab:main-results}, we show the metric values when we regard the original image and the ground truth image as the model output. If the models succeeded in editing the images according to the editing prompt, the metric averages for the LLM/LMMs should have improved than the ``Original'' values. Also, we report the percentage of images that caused an error during the evaluation pipeline in the rows on the right. Processing failures occur in two cases: when our pipeline cannot extract the SVG code from the LLM/LMM output and when rasterization fails due to grammatical errors in the SVG code or other reasons. In \Tref{tab:main-results}, we report the former case as ``Extraction'' and the latter as ``Parsing''.

\subsection{Qualitative analysis}
We can draw the following conclusions from the metrics and the error rates in \Tref{tab:main-results}.
\begin{enumerate}
  \item In general, \textbf{it is difficult for the current LLMs/LMMs to solve image editing tasks that require a semantic understanding of the images}. We can see that most averages are worse than the ``Original''. These values suggest that the editing models made the images farther from the ground truth.
  \item There was no clear tendency indicating the effect of the vision capabilities of LMMs with SVG editing (\eg, Mistral v0.2 MSE: $0.115$ vs. LlaVa-NeXT MSE: $0.110$, Mistral v0.2 CLIPScore: $25.474$ vs. LlaVa-NeXT CLIPScore: $25.401$). Finely adjusting the parameters is crucial when editing SVG images, and only having vision capabilities is insufficient to improve the performance.
  \item By training LLMs on code, the models could generate SVG code more accurately, improving the error rate (\eg, Llama 2 Extraction: $33.1\%$ vs. Code Llama Extraction: $14.1\%$). This result suggests that having SVG code as training data helps precisely generate SVG code. However, it does not seem to impact the evaluation metrics significantly (\eg, Llama 2 MSE: $0.102$ vs. Code Llama MSE: $0.103$).
  \item The ability to generate SVG code correctly depends on the model series. For instance, the failure rate exceeds 60\% for Phi-3.5 while it is under 10\% for Mistral. This trend is more apparent between Open and Closed models; the error rate is near zero for most Closed models.

\end{enumerate}

\subsection{Quantitative analysis}
In this section, we will look closer at the editing results and clarify when and how the current LLMs/LMMs succeed or fail to help future research on SVG editing.

There were instances where the models could output images similar to the ground truth for some relatively simple tasks. We can classify the successful cases into two categories: deleting and applying basic transformations to elements. In the transforming tasks, the LLMs could properly utilize functionalities of SVG (\texttt{transform} attribute \etc). Also, the LLM/LMM could sometimes accurately calculate the coordinates of the control points, taking the entire drawing range into account. These examples show that current LLMs/LMMs understand basic SVG grammar and can recognize the meaning of each object from the code.

Looking at the LLM/LMM outputs that the evaluation pipeline regarded as an error, we found these errors prominent in the \texttt{<path>} element. In some examples, the language models repeated similar number sequences, resulting in an incomplete code. The \texttt{<path>} elements often include a complex array of numbers. In other cases, the \texttt{d} attribute of an output \texttt{<path>} element contained undefined line commands, or the number of arguments was too many or insufficient.

LLMs/LMMs also seemed likely to avoid or fail processing \texttt{<path>} elements. In some cases, the model had only added an \texttt{<text>} element when it had to reform a shape. Even when the model attempted to edit \texttt{<path>} elements, the output had messy and corrupted shapes.
\section{CONCLUSION}
We developed a benchmark to evaluate the performance of instruction-based editing of SVG images, SVGEditBench V2. We used LLMs and LMMs as SVG editing techniques and conducted comparative experiments. The results show that they understood the basic grammar of SVG and could sometimes understand the images through code. In those cases, they could successfully edit the graphics according to the text prompt. However, the experiments indicated that the editing ability was far from satisfactory. Qualitative analysis showed that the main issue lies in generating or editing \texttt{<path>} elements since elaborate control of numbers is critical.

Considering these conclusions, future SVG editing models should focus on fine-grained control of numbers. Language models for SVG processing might need an assistive module for error-free and high-quality path generation. Vector graphics are abundant in real-world situations, and research on text-to-vector is desirable. We hope SVGEditBench V2 facilitates research in this area and helps ease vector graphics processing for both professionals and novices.

\bibliographystyle{IEEEbib}
\bibliography{main}
\end{document}


\maketitle
\section{DATASET DOCUMENTATION}
\subsection{Intended use}
We introduced SVGEditBench V2, a dataset and benchmark for SVG editing. The benchmark aims to evaluate editing models that take an image and an editing prompt as input and output the image modified in accordance with the prompt. Training an editing model was not our primary focus when building the dataset. However, the dataset creation pipeline can easily be extended in producing editing datasets on a larger scale, \ie~for training. The pipeline can also create datasets for different types of SVG images (not just emojis).

\subsection{Availability of the dataset and code}
The code and data related to this work are available and will be preserved at \url{https://github.com/mti-lab/SVGEditBenchV2}. This GitHub repository includes the code for creating the benchmark and evaluating LLMs/LMMs. The repository also hosts the built dataset except for the images. The repository contains the code for restoring the images into the dataset.

The license for using the images to create the dataset is as follows.
\begin{itemize}
  \item Noto Emoji: Apache license, version 2.0
  \item Twemoji: CC BY 4.0 license
  \item Fluent Emoji: MIT License
\end{itemize}
The code is licensed under the MIT license. The authors bear the responsibility in case any issues related to data license arise.
\section{DETAILS OF THE DATASET CREATION PIPELINE}
This section explains the dataset creation pipeline in more detail than the main paper.

\subsection{Method of retrieving descriptions of emojis}
\label{subsec:emoji-names}
As we explained in Sec. 3.1, emojis have a name associated with them. We use this information in both dataset creation and evaluation. Please refer to Sec. 3.2 and Sec. 3.6 for how we used the descriptions in each pipeline.

The following algorithms show how we gained the descriptions of the images used in our dataset:

For images in Fluent Emoji, the SVG image is under the folder with a descriptive name. Therefore, we use this folder name as the description.

For images in Twemoji and Noto Emoji, the filename of the SVG image indicates the associated Unicode codepoints. We use the following steps to obtain the description from these codepoints:
\begin{enumerate}
    \item If the number of codepoints used to express the emoji is one, we use the \texttt{unicodedata} library ~\cite{unicodedata} in Python 3.12 to retrieve the emoji name. This library can retrieve the character properties in the Unicode Character Database version 15.0.0. This step returns an error if the emoji is in the Private Use Area. For instance, Twemoji contains an emoji in the Private Use Area: Shibuya109 (U+E50A). If this happens, we exclude the image from the pipeline.
    \item If the number of codepoints is more than two, we first search it on the list of ZWJ sequences available on the \textit{Unicode, Inc.} website~\cite{zwj-sequences}. In ZWJ sequences, a single glyph uses multiple emojis with a ZERO WIDTH JOINER (U+200D) in between. If the sequence of codepoints is present in the list, we use the description in the list.
    \item If the sequence is not on the list, we look at the final codepoint. If it is \texttt{U+20E3}, the emoji is a keycap (\ie, a number or a symbol on a square with rounded corners). We run the previous steps except for the final codepoint and prepend \texttt{Keycap:} to the returned description. If the last codepoint is \texttt{U+FE0F}, we simply run the steps without the final codepoint. We can do this because \texttt{U+FE0F} is called the ``emoji variation selector'' and does not change the meaning of the emoji.
    \item If we cannot obtain the description with these steps, we regard it as an unknown ZWJ sequence. We split the sequence with \texttt{U+200D} and get the description for each fragment. We concatenate all the descriptions with a \texttt{+} sign between them and use this to describe the image as a whole.
\end{enumerate}

\subsection{Details of image pair extraction}
\label{subsec:image-pair-extraction}
\begin{figure*}
  \begin{minipage}{0.33\textwidth}
    \includegraphics[width=\linewidth]{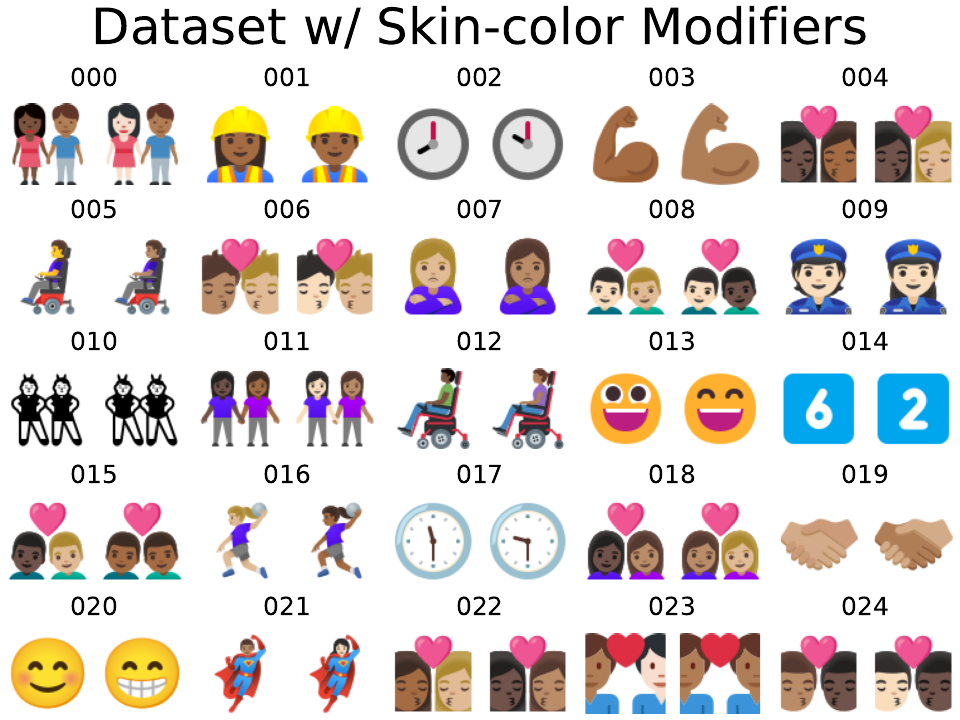}
    \caption{Examples of image pairs created with emojis with modified skin color included. The tasks are biased to tasks involving changing color.}
    \label{fig:ablation-skin-color}    
  \end{minipage}
  \hspace{0.03\linewidth}
  \begin{minipage}{0.63\linewidth}
      \includegraphics[width=0.50\linewidth]{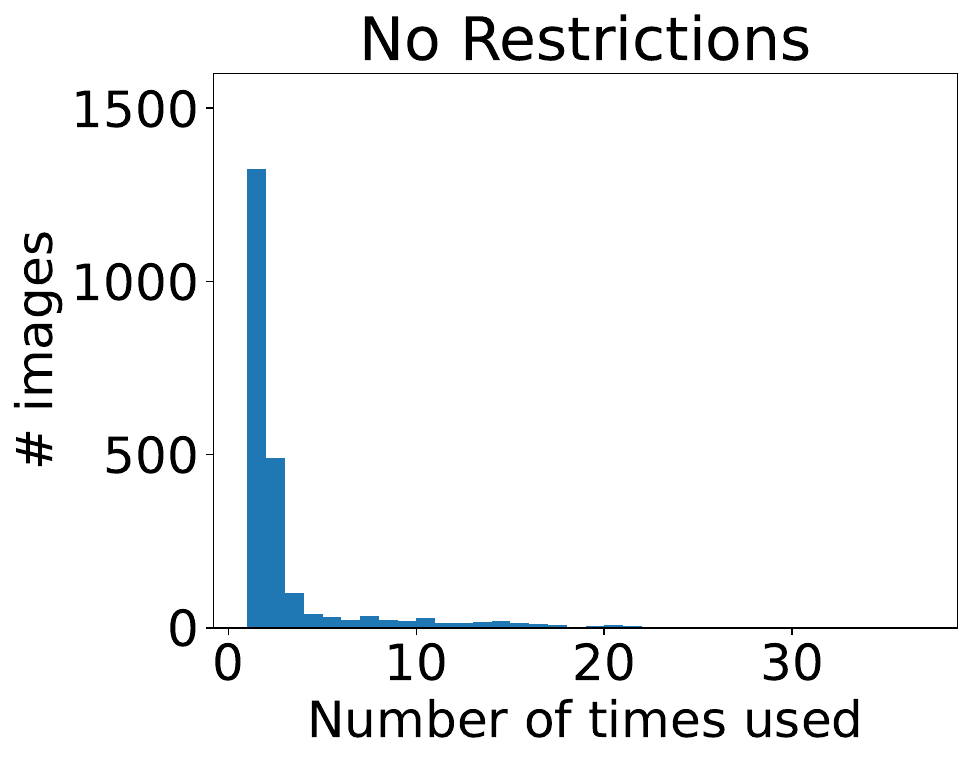}
      \includegraphics[width=0.50\linewidth]{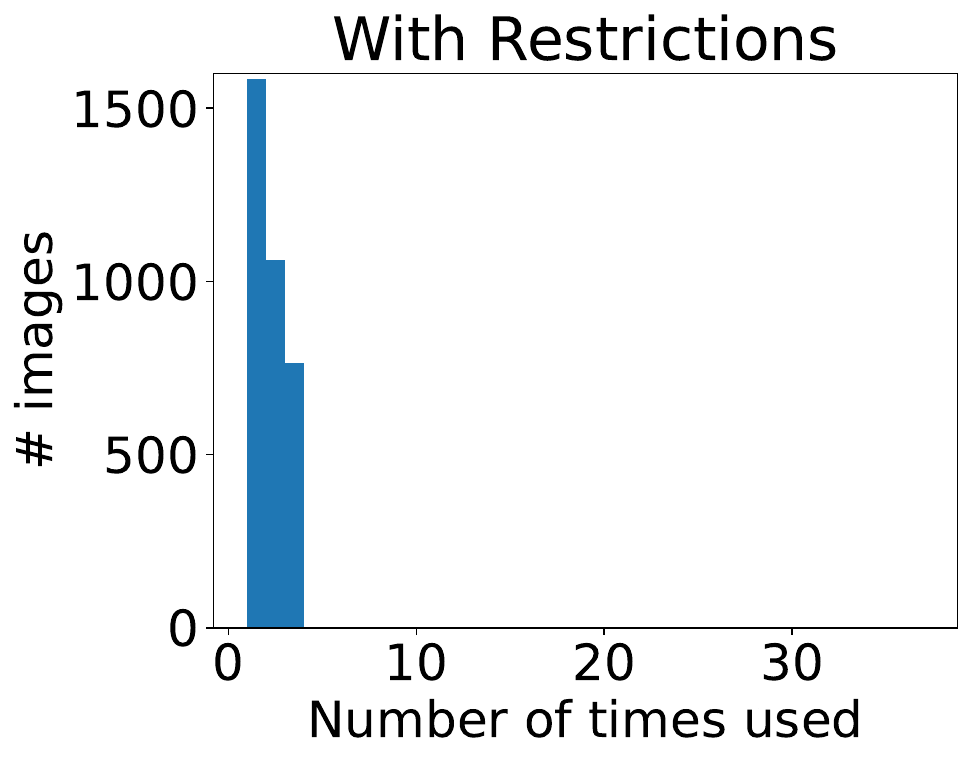}
      \caption{The distribution of the number of times the same image appears in the dataset. The left histogram shows the distribution when we applied no restrictions, and the right shows the distribution in our proposed dataset (\ie, the same image can be in the dataset only up to three times).}
      \label{fig:ablation-restriction}
  \end{minipage}
\end{figure*}

Before extracting image pairs, we excluded the images that seemed inappropriate for the editing tasks. Specifically, we did not use flag images and emojis whose codepoints have skin color modifiers (U+1F3FB to U+1F3FF). The reason is that incorrectly editing these images may offend some people. 

Ensuring the diversity of tasks is another reason for excluding images with modified skin color. \Fref{fig:ablation-skin-color} shows the example image pairs built when those emojis are present. This figure reveals that the dataset contains a lot of Change Color tasks if we used emojis with those emoji modifiers. This trend is because the image pairs with only differences in skin color will be close in structure and semantics. Our distance metric used here regards these pairs as near. Note that some Change Color tasks remain even if we exclude such kinds of emojis (e.g., the middle-right example of Fig. 1).

We extracted the image pairs using the following algorithm:
\begin{enumerate}
    \item We calculate the distance metric discussed in Sec. 3.2 for all combinations of two images in the datasets. Since the images are SVG, we rasterized them at $64\times 64$ resolution before calculating LPIPS.
    \item We sort the combinations with the metric so that the pair closer in distance is processed earlier.
    \item If the metric value is under 0.1, we exclude the image pairs. This step disregards pairs with identical images. Pairs of identical images occur partly because we used two versions for the Fluent Emoji datasets: the flat and high-contrast versions. The emojis in the high-contrast version are black-and-white variants of the flat ones. Therefore, if those in the flat version already include only black and white (\eg, club or spade suits), the corresponding high-contrast version will be identical to the flat version.
    \item If either of the images is already used more than three times, we skip the image pair.
    \item We randomly choose which image is the original and which is the ground truth for each pair.
    \item We end the extraction process if we obtain 3000 image pairs.
\end{enumerate}

\begin{figure*}
  \centering
  \subfloat[LPIPS only]{
    \includegraphics[width=0.3\linewidth]{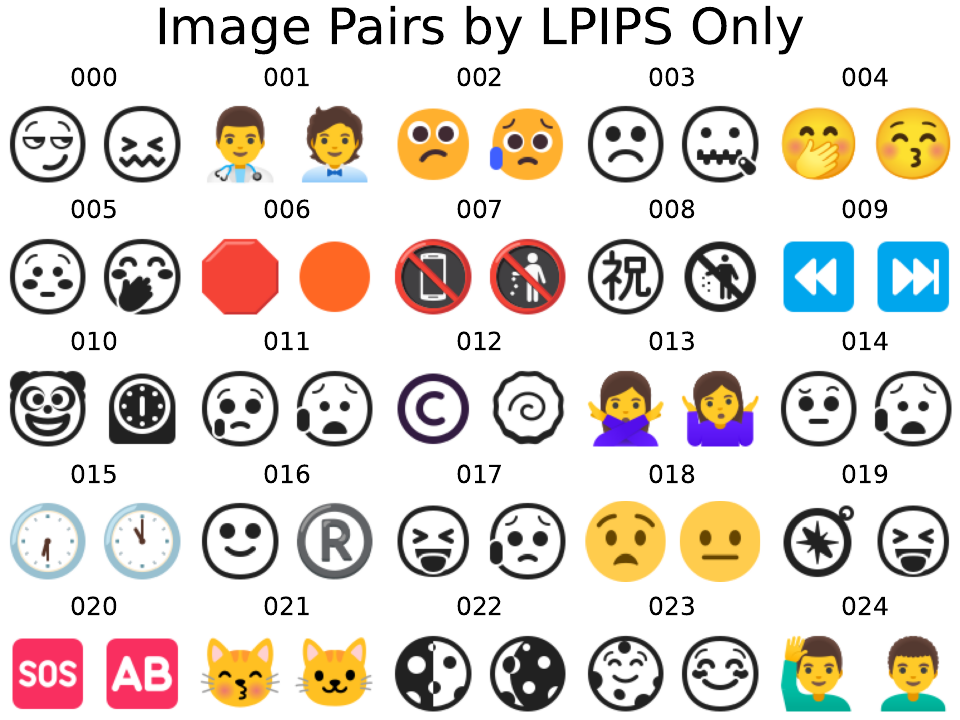}
    \label{fig:ablation-creation-metrics-LPIPS}
  }
  \hspace{0.03\linewidth}
  \subfloat[CLIP similarity only]{
    \includegraphics[width=0.3\linewidth]{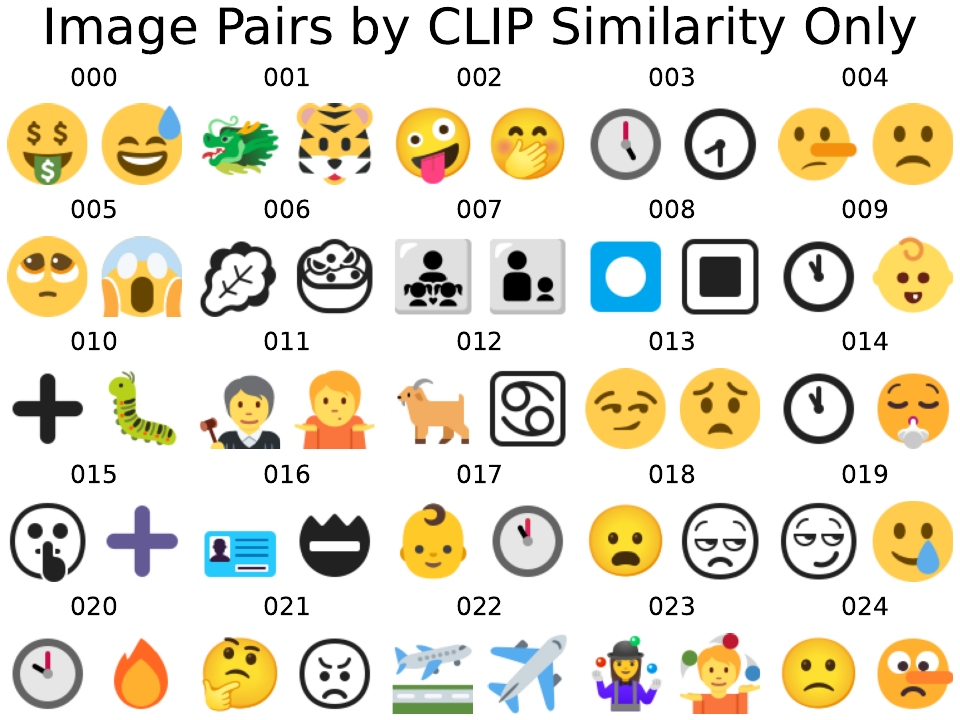}
    \label{fig:ablation-creation-metrics-CLIP}
  }
  \hspace{0.03\linewidth}
  \subfloat[LPIPS + CLIP similarity]{
    \includegraphics[width=0.3\linewidth]{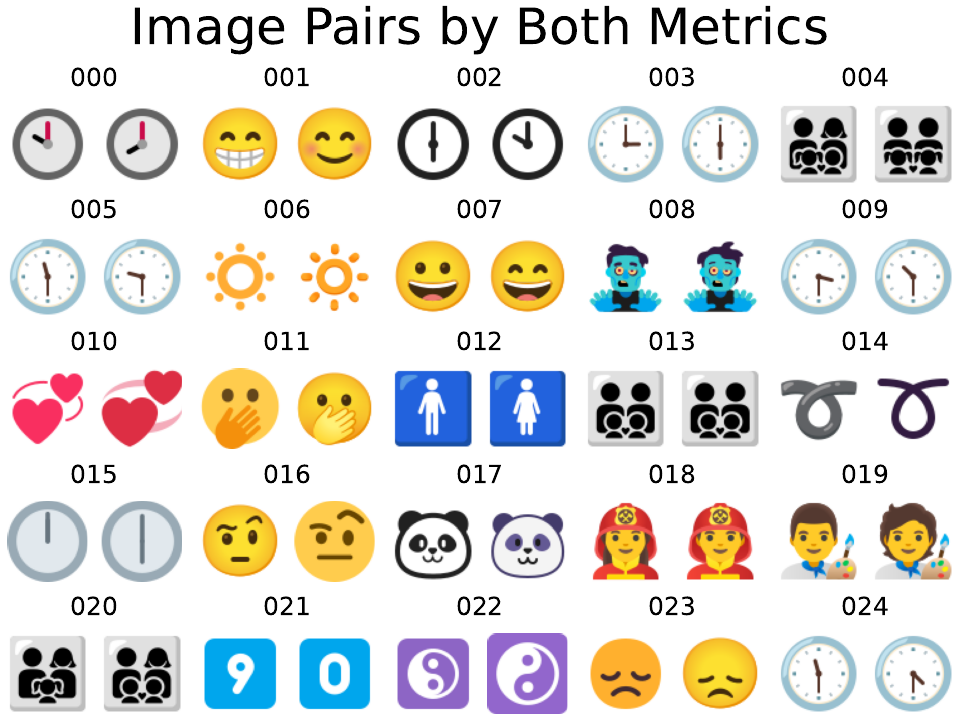}
    \label{fig:ablation-creation-metrics-both}
  }
  \caption{Examples of image pairs extracted using different comparison metrics. The numbers above the image pairs are assigned in order of the distance metric (closer to farther). The image pairs on the left look similar but differ semantically, while the opposite is true in the middle pairs. Using both image- and text-based metrics leads to image pairs with realistic and easily describable editing tasks.}
  \label{fig:ablation-creation-metrics}
\end{figure*}

We show an ablation study on the effectiveness of restricting the frequency at which the same image can appear (Step 4 in the above algorithm). We introduced this step to ensure the variety of the emojis. \Fref{fig:ablation-restriction} shows the distribution of how many times a single image is in the dataset with and without the restriction. We see that the dataset without this restriction includes some emojis many times, 36 times at maximum. In addition, we could consist of 52\% more images (2,246 without the limit and 3,409 with the limit) by setting this limit.

Additionally, we show how the image comparison metric contributes to gaining appropriate image pairs. We used the sum of LPIPS of the rasterized images and the cosine distance of CLIP text embeddings of the descriptions in our dataset creation pipeline. We aim to extract image pairs that are similar both visually and semantically by using this metric. This characteristic is vital since we believe editing occurs between images similar in both appearance and meaning.

\Fref{fig:ablation-creation-metrics} shows some image pairs extracted using different metrics. We obtained \Fref{fig:ablation-creation-metrics}\subref{fig:ablation-creation-metrics-LPIPS} and \Fref{fig:ablation-creation-metrics}\subref{fig:ablation-creation-metrics-CLIP} solely with LPIPS and CLIP text distance, respectively. \Fref{fig:ablation-creation-metrics}\subref{fig:ablation-creation-metrics-both} is from the generated dataset before filtering (thus, it may contain pairs not used in experiments discussed in Sec. 4). The image pairs in \Fref{fig:ablation-creation-metrics}\subref{fig:ablation-creation-metrics-LPIPS} look similar in shapes and colors. However, the meanings of the images may not be related. For example, the images in pair 008 in \Fref{fig:ablation-creation-metrics}\subref{fig:ablation-creation-metrics-LPIPS} mean ``congratulations'' and ``no littering,'' respectively. Conversely, the image pairs in \Fref{fig:ablation-creation-metrics}\subref{fig:ablation-creation-metrics-CLIP} express similar concepts (\eg, images in pair 016 represent ``identification card'' and ``name badge'', respectively). However, the images may look totally different. Differences in these image pairs will be too complex to describe concisely in text. We obtain image pairs that look and mean similar by summing them, as in \Fref{fig:ablation-creation-metrics}\subref{fig:ablation-creation-metrics-both}.

\subsection{Prompt for Retrieving Editing Instructions}
\label{subsec:appendix-instruction-prompt}
We provided GPT-4o with the following prompt to generate the editing instructions. The rasterized versions of the original and ground truth images come after this text prompt.
\begin{itembox}[l]{The prompt for generating editing instrucions}
\begin{lstlisting}
The first image is an emoji of "<b#(Caption of the original image)#>" and the second is an emoji of "<b#(Caption of the ground truth image)#>." Describe how the first image should be edited to look like the second image. Do not just say "Change to match the second image/emoji," but specify the the expected result. Also, make the instruction as clear and as short as possible.

For example, if a plane is landing towards the runway in the first image and taking off in the second, you could say "Make the plane take off."
\end{lstlisting}
\end{itembox}

The first sentence states what the subsequent two images represent. We used the descriptions obtained as explained in \Sref{subsec:emoji-names} for the captions. The second sentence asks GPT-4o to create the editing instruction. The following two sentences suggest that the editing instruction should not mention the second image (ground truth image), and GPT-4o should provide clear and concise instructions. The instruction should avoid referencing the second image because the target SVG editing model will not see the ground truth image while editing. The final sentence shows an example of captions and their instruction. Showing this example aims to ensure variety in the expression of the prompts (\ie, not just ``Change \textit{A} into \textit{B}'').

\begin{figure}
    \centering
    \includegraphics[width=\linewidth]{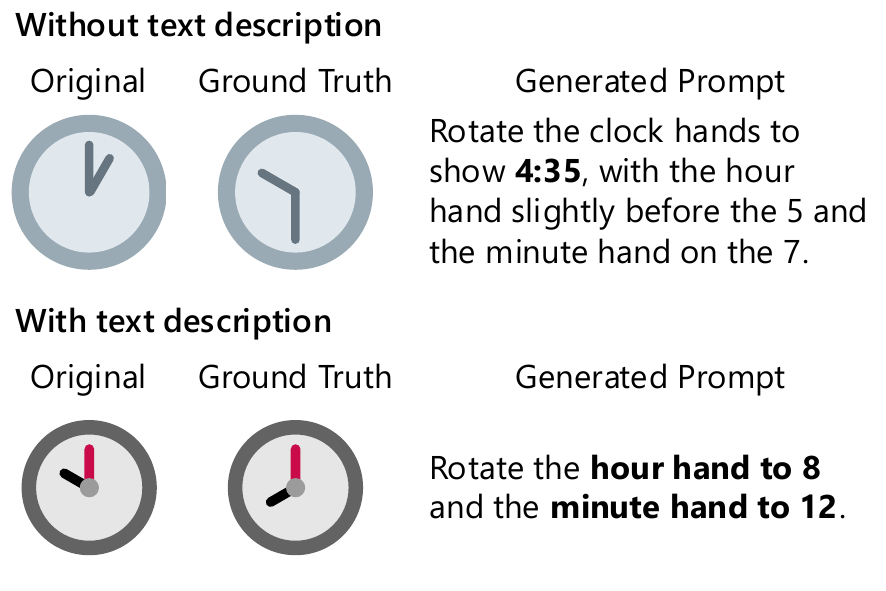}
    \caption{Example of image captions helping generation of precise editing instructions. The time mentioned in the prompt is wrong in the top example, while the instruction is accurate in the bottom.}
    \label{fig:clock}
\end{figure}
The biggest motivation for using captions in this prompt is that GPT-4o occasionally cannot understand the image's content accurately, especially for clocks. \Fref{fig:clock} shows one example of how these captions contribute to generating precise prompts. In this example, GPT-4o recognized the time in the ground truth image as 4:35 instead of 10:30 when it did not see the caption. In contrast, it could accurately describe where the clock hands should be when considering the two images' captions.

\subsection{Details of the Filtering Stage}
\begin{figure}
    \centering
    \includegraphics[width=\linewidth]{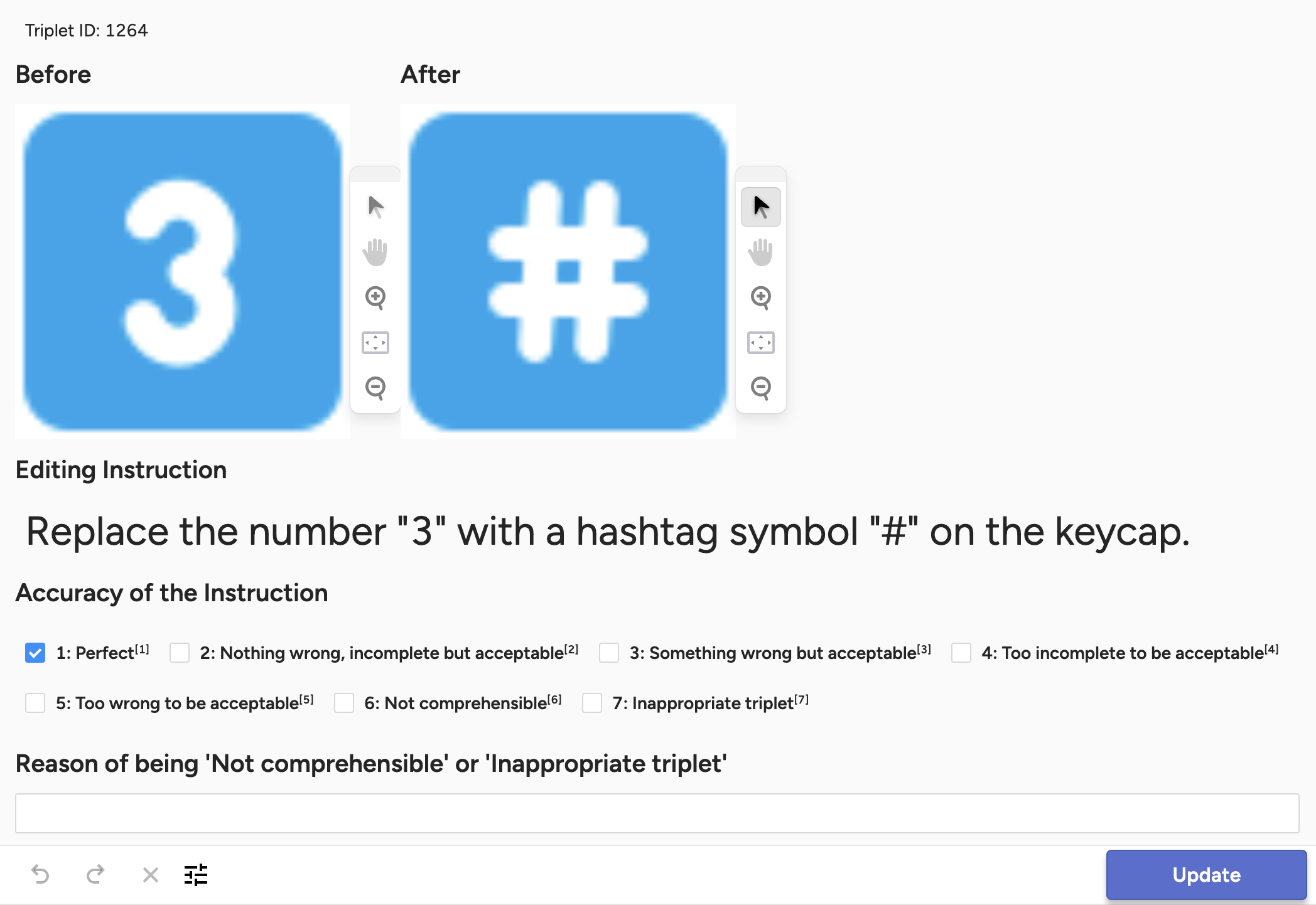}
    \caption{The interface for filtering the triplets. We used Label Studio to filter out inaccurate editing prompts or inappropriate images.}
    \label{fig:label-studio}
\end{figure}
We manually filtered the generated triplets since image pairs contain inappropriate images or GPT-4o may produce inaccurate prompts. This section looks into the filtering procedure in detail. 

We looked at the image pairs and subjectively classified all the triplets into the following seven categories. We included triplets classified as 1., 2., or 3. in our dataset. We used Label Studio~\cite{labelstudio} to label the triplets smoothly. \Fref{fig:label-studio} shows the interface we created on Label Studio.

\begin{enumerate}
    \item Perfect: The generated editing prompt mentions every component different between the original and ground truth images.
    \item Nothing wrong but not complete: The prompt contains no false statements but does not mention some subtle components different between the two images.
    \item Something wrong but acceptable: The prompt contains slight errors that seemingly do not affect the output.
    \item Too incomplete to be acceptable: The prompt does not mention the main change between the images.
    \item Too wrong to be acceptable: The prompt has critical false statements.
    \item Not comprehensible: The prompt is not in proper English.
    \item Inappropriate triplet: The content of the images seems inappropriate to include in the dataset.
\end{enumerate}

\begin{figure*}
    \centering
    \includegraphics[width=\linewidth]{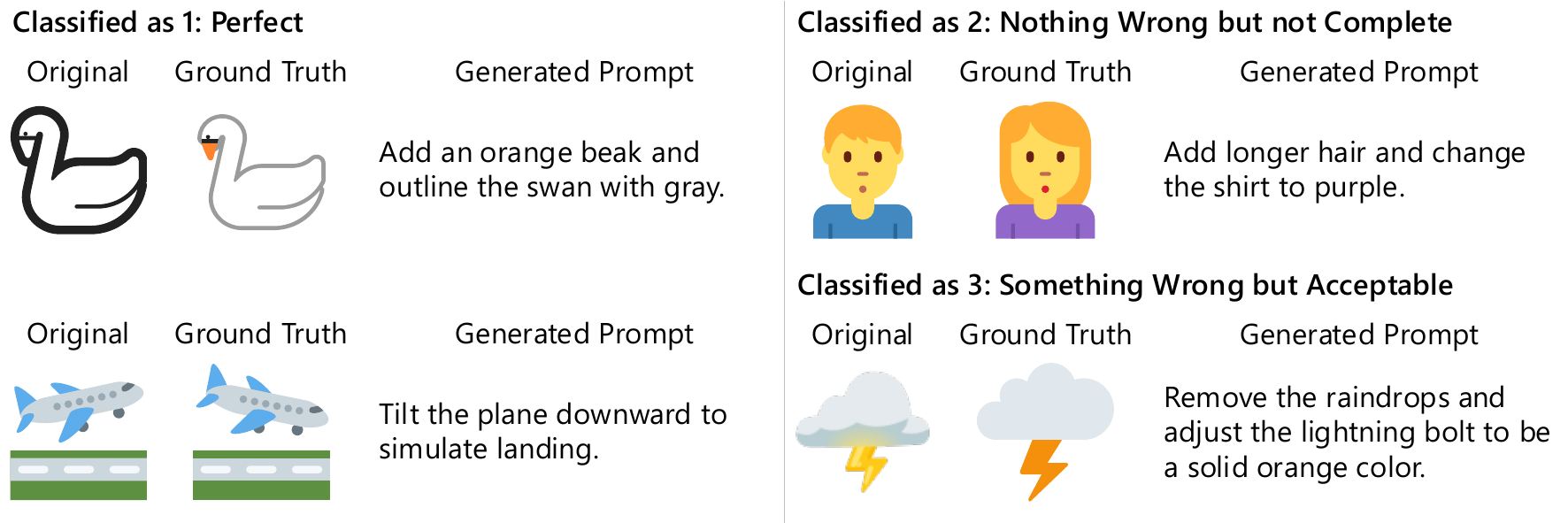}
    \caption{The example triplets included in our dataset}
    \label{fig:filtering-success}
\end{figure*}
\begin{figure*}
    \centering
    \includegraphics[width=\linewidth]{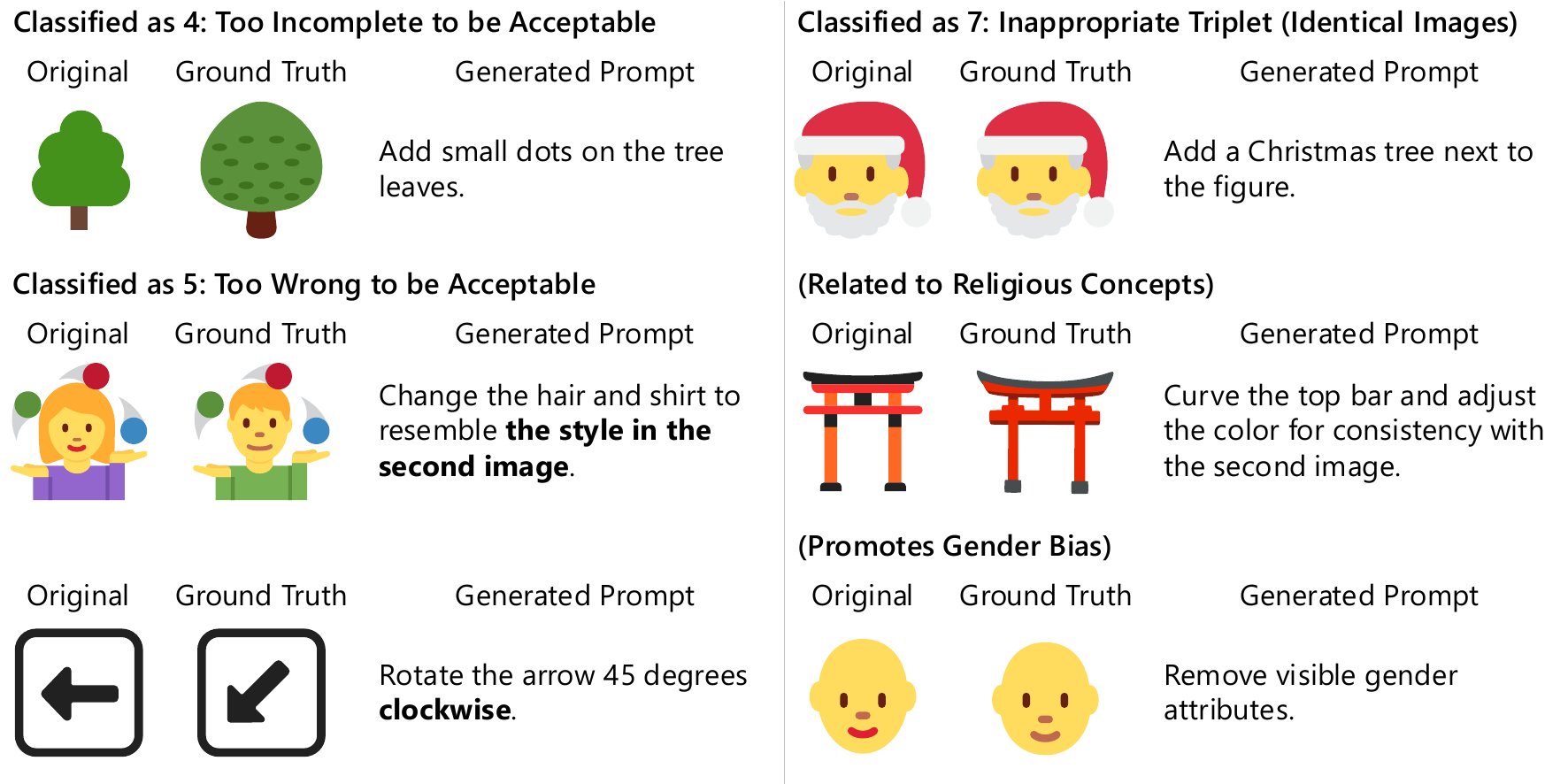}
    \caption{Some notable examples excluded in the filtering process}
    \label{fig:filtering-failure}
\end{figure*}
We show some example triplets in each category to show interesting prompts generated by GPT-4o. \Fref{fig:filtering-success} shows some successful triplets, and \Fref{fig:filtering-failure} shows some failure cases. No triplets fell into Category 6. 

We first describe the successful cases and why we classified those triplets into each category. The top-left of \Fref{fig:filtering-success} shows an example of GPT-4o looking at the images and accurately telling the differences between them, not just by comparing the prompts. The two images have the same caption: ``swan''. However, the prompt generated by GPT-4o could tell that the outline in the ground truth image is gray and the beak is orange. The bottom-left triplet shows how the example in the prompt (see \Sref{subsec:appendix-instruction-prompt}) affected the results. The image pair of this example is just the opposite of the example in the prompt. Therefore, one may expect the generated prompt to be ``Make the plane land.'' However, GPT-4o could invent a different way of telling the same thing. The top-right is one of the triplets classified as ``Nothing wrong but not complete.'' The generated prompt here could indicate that the hair and shirt color differ, but it could not tell that the mouth color changed to red. However, we included this in our dataset since the mouth does not cover much of the image, and the prompt could mention the main differences between the emojis. The bottom-right example was classified as ``Something wrong but acceptable.'' The prompt in this triplet mentioned the raindrops that are not present in the original image. We included this triplet because removing the raindrops is impossible, meaning complying with this part would not change the outcome.

We'll now describe the failure cases in \Fref{fig:filtering-failure}. The top-left is one example of a triplet in Category 4. The editing instruction could not pinpoint the main difference: the tree's shape. The bottom two examples in the left column are from Category 5: Too wrong to be acceptable. One notable example here is that the prompt mentioned the second image, although we asked not to do so in the prompt (see \Sref{subsec:appendix-instruction-prompt}). We excluded such triplets since the editing model could not know how to edit the image. Another type of failure is that GPT-4o could be wrong in visual concepts, as in the bottom-left example. It instructs to turn the arrow clockwise, while it should say counter-clockwise. In another generated prompt, GPT-4o mistook gray for light blue, which would significantly affect the editing result. Therefore, we excluded these triplets from our dataset.

There were four types of inappropriate triplets (Category 7), and we illustrated three of them in the right row of \Fref{fig:filtering-failure}. 
\begin{enumerate}
    \item Some image pairs had identical original and ground truth images. These two characters had different Unicode codepoints and, thus, different captions. The original image was ``FATHER CHRISTMAS'' (U+1F385), while the ground truth image was captioned as ``MAN + CHRISTMAS TREE'' (U+1F468, U+200D, U+1F384). Since the images have different captions, the distance metric explained in \Sref{subsec:image-pair-extraction} was big enough not to be excluded in Step 1. The prompt also suggests that the captions of the images greatly influenced the generated instructions.
    \item Some images were strongly related to religious concepts, as in the middle row of \Fref{fig:filtering-failure}. We considered these images unsuitable for our dataset for the same reason for excluding flags. 
    \item We considered the bottom-right example as promoting gender bias in the evaluation. To successfully edit the original image, the editing model should recognize what the ``gender attributes'' point to. 
    \item The emoji datasets included images that directly mean something offending, such as the ``REVERSED HAND WITH MIDDLE FINGER EXTENDED'' (U+1F595). We excluded triplets with those images.
\end{enumerate}

\subsection{Additional analysis of the created dataset}
\begin{figure*}
  \centering
  \begin{minipage}{0.57\hsize}
      \centering
      \includegraphics[height=5cm]{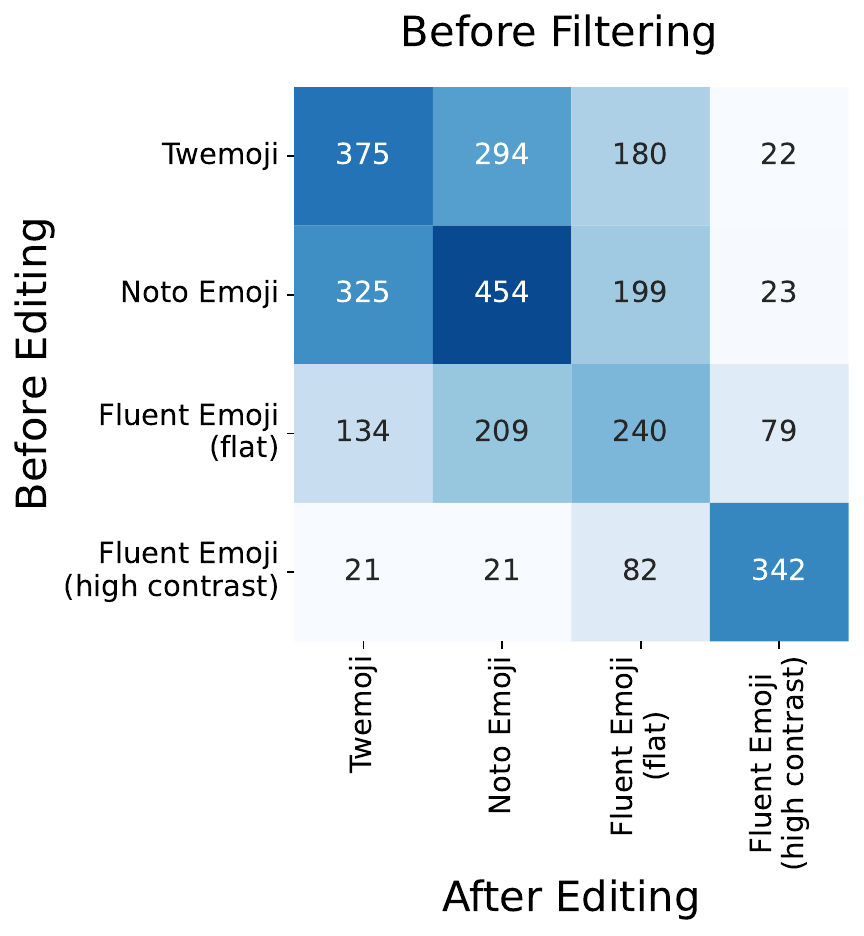}
      \includegraphics[height=5cm]{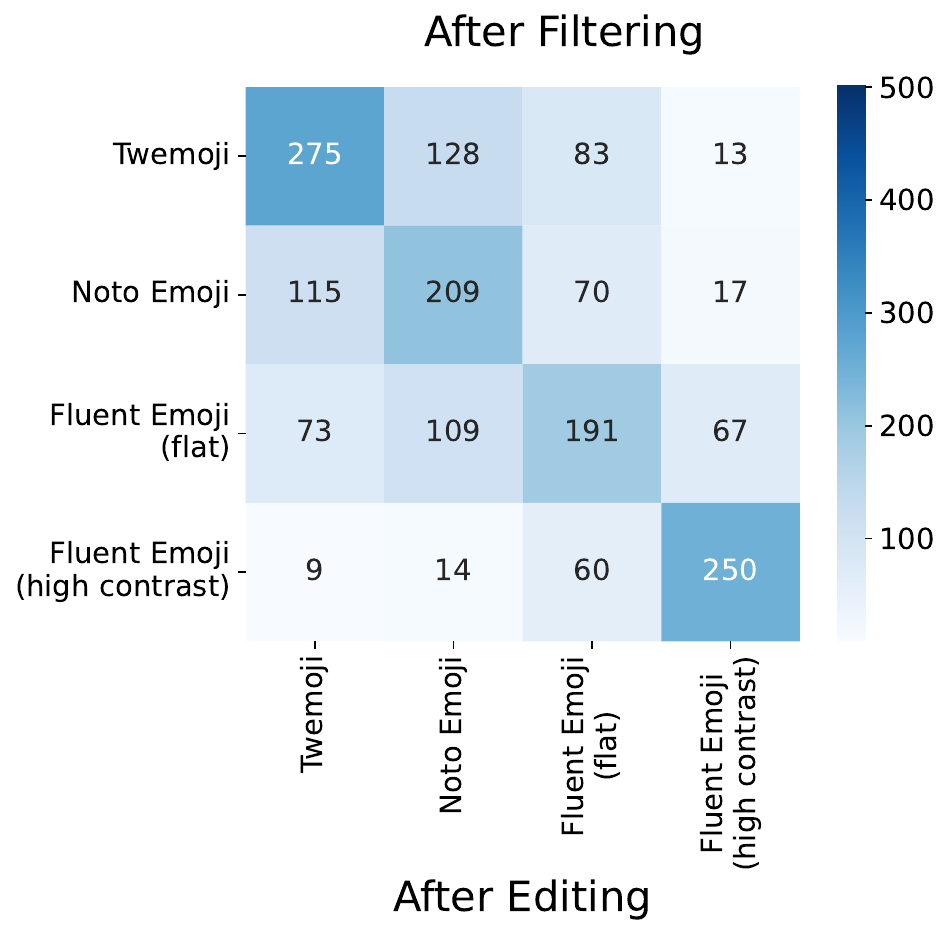}
      \caption{Distribution of the SVG emoji datasets to which the images in the created dataset belong. The left figure shows the distribution of triplets before filtering, and the right shows the distribution after filtering.}
      \label{fig:distribution}
  \end{minipage}
  \hspace{1em}
  \begin{minipage}{0.4\hsize}
    \centering
    \includegraphics[height=4cm]{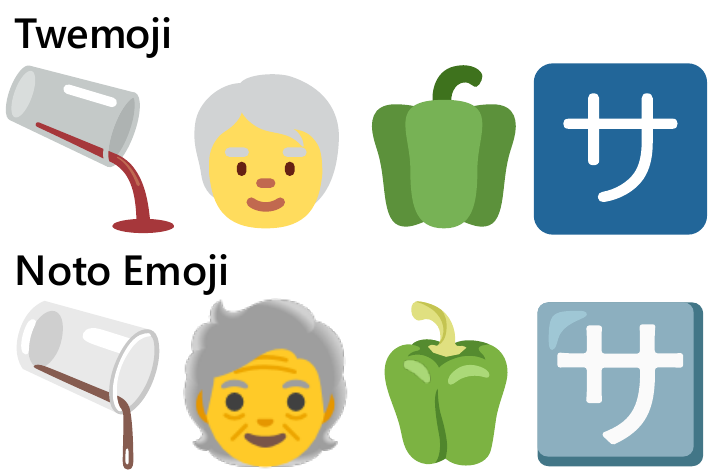}
    \caption{Images from Twemoji and their corresponding images from Noto Emoji. Emojis in Twemoji are mostly flat and simpler, while emojis in Noto Emoji tend to be more detailed. These image pairs are the ones actually included in the dataset.}
    \label{fig:dataset-comparison}
  \end{minipage}
\end{figure*}
This section shows the statistics of the dataset. We focus on which dataset the images in the generated dataset originated. \Fref{fig:distribution} shows the distribution of the origin dataset. The figure shows that the images before and after editing mainly come from the same datasets, but image pairs from distinct datasets are also present.

This result verifies the effectiveness of using multiple datasets as the source data. Emojis have a unique style depending on the dataset to which they belong. We show the style differences between datasets in \Fref{fig:dataset-comparison}. In this figure, the emojis in each column have the same Unicode codepoint. The above images are from Twemoji, and the bottom are from Noto Emoji. These image pairs are in the built dataset. We could obtain such style transfer tasks only by using multiple source datasets. 
\section{DETAILS ON MODEL EVALUATION}
This section explains additional details of the model evaluation pipeline that were not in the main paper.
\subsection{Prompt Used for Inferencing LLMs for SVG Editing}
\label{subsec:prompt-evaluation}
We used the following prompt to infer the LLMs/LMMs when performing SVG editing. We kept this simple, aiming to evaluate the pure ability of LLMs/LMMs to process SVG. 

\begin{itembox}[l]{The prompt used for evaluating LLMs/LMMs}
\begin{lstlisting}
The following is the SVG code representing an image. <b#(Instruction)#> Only return the output SVG code.

```svg
<b#(SVG code before editing)#>
```
\end{lstlisting}
\end{itembox}

\subsection{Extracting SVG code from the LLM output}
\label{subsec:code-extraction}
Using the following algorithm, we extracted the SVG code from the LLM/LMM output. Note that we did not look into the content of the code at this step.
\begin{enumerate}
    \item We look for parts surrounded with \texttt{```svg} and \texttt{```}. We regard the output as an ``Extraction'' error if there are multiple code segments. If there is only one segment, we regard this code segment (excluding \texttt{```svg} and \texttt{```}) as the output SVG code.
    \item If there are no code segments in the previous step, we look for text surrounded with \texttt{<svg} and \texttt{</svg>}. We consider the text (including \texttt{<svg} and \texttt{</svg>}) as the SVG output if there is only one segment and as an ``Extraction'' error if otherwise.
\end{enumerate}

\subsection{Definition of the two-step Chamfer distance}
The following is the definition of the Chamfer distance metric used in the model evaluation pipeline. Let $X$ be the output image. We assume $X$ as a set of shapes $A$. Therefore $X = \{A_1, A_2, \dots\}$. Also, we assume each shape $A$ as a set of two-dimensional points. Therefore $A=\{\bm{a}_1,\bm{a}_2, \dots \}$, where $\bm{a} \in \mathbb{R}^2$. We obtain these points by equally sampling the contour of each shape ($|A|=100$). The ground truth image $Y=\{B_1, B_2, \dots\}$ can be defined similarly. Given this, the distance between $X$ and $Y$ is defined as follows. Firstly, we define the distance between shapes as 
\newcommand{\mshape}{\mathrm{shape}}
\begin{align}
\begin{autobreak}
    d_\mshape\left(A, B \right)
    = \frac{1}{|A|}\sum_{\bm{a} \in A}\min_{\bm{b} \in B}\left\| \bm{a} - \bm{b} \right \|_2^2 
    + \frac{1}{|B|}\sum_{\bm{b} \in B}\min_{\bm{a} \in A}\left\|\bm{a} - \bm{b}\right\|_2^2 .
\end{autobreak}
\end{align}
Then, we define the distance between the two images as
\begin{align}
\begin{autobreak}
     d_\mathrm{image}\left(X, Y\right)
     = \frac{1}{|X|}\sum_{A \in X}\min_{B\in Y}d_\mshape\left(A, B\right) 
     + \frac{1}{|Y|}\sum_{B \in Y}\min_{A \in X} d_\mshape\left(A, B\right).
\end{autobreak}
\end{align}

\section{EXAMPLE OUTPUTS IN THE EXPERIMENT}
\begin{figure*}
  \includegraphics[width=\hsize]{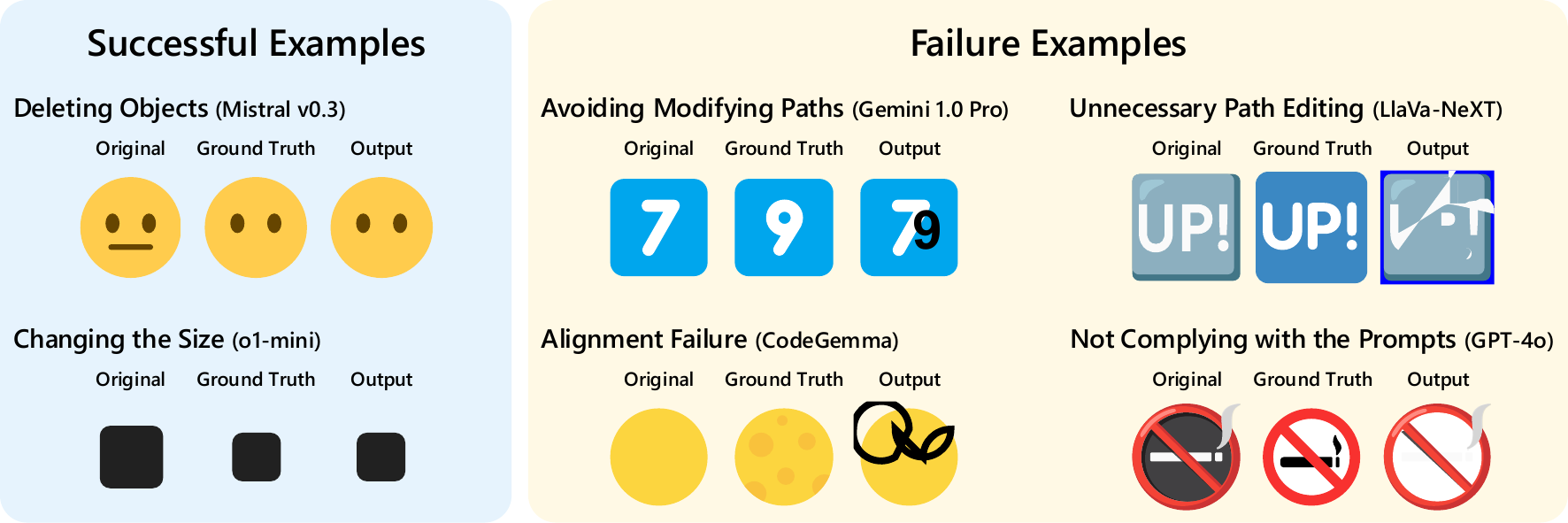}
  \caption{Examples of current LLMs and LMMs performing the editing tasks in SVGEditBench V2. The first column shows successful examples, and the two columns on the right show failure cases.}
  \label{fig:examples}
\end{figure*}

We show some samples of actual SVG output from the LLMs/LMMs in the experiment. The two instances on the left of \Fref{fig:examples} are from the successful cases mentioned in Sec. 4.4. We introduce more patterns of failures in the subsequent sections.

\subsection{Cases of ``Extraction'' errors}
\newsavebox\repetitionbox
\newsavebox\noclosetagbox
\newsavebox\ignoringbox

\begin{lrbox}{\repetitionbox}
  \begin{minipage}{0.96\linewidth}
    \begin{lstlisting}
```svg
<?xml version="1.0" encoding="utf-8"?>
<b#...#>
<path style="fill:#fff" d="M10,10l5,5 5,-5 5,<b# ... #>5 5,-5 5,5 5,-5 5,5 5,-5 5,5 5,-5 5,5 5,-5 5,5 5,-5 5,5 5,-5 5,5 5,-5 5,5 5,-5 5,5 5,-5 5,5 5,-5 5,5 5,-5 5,\end{lstlisting} 
    \end{minipage}
\end{lrbox}

\begin{lrbox}{\noclosetagbox}
  \begin{minipage}{0.96\linewidth}
    \begin{lstlisting}
<svg width="32" height="32" viewBox="0 0 32 32" fill="none" xmlns="http://www.w3.org/2000/svg">
<b#...#>
<path d="M14.6131 7.80203C15.5509 8.03583 16.1215 8.9856 15.8877 <b#...#> 13.6753 7.56822 14.6131 7.80203Z" fill="#FFFFFF"/>\end{lstlisting} 
  \end{minipage}
\end{lrbox}

\begin{lrbox}{\ignoringbox}
  \begin{minipage}{2\linewidth}
  \begin{lstlisting}
To add a sandy texture to the umbrella's base, we need to modify the "fill" attribute of the base path element (the fourth path in your SVG code). Instead of a solid color, we will use a pattern to create the sandy texture.

First, let's create an SVG pattern element for the sandy texture. We'll use two colors (light brown and darker brown) and some noise to simulate sand.

```svg<b#...#>```

Now, update the base path to use the sandy-texture pattern:

```svg<b#...#>
  \end{lstlisting} 
\end{minipage}
\end{lrbox}

\begin{figure*}
  \centering
  \subfloat[An example of repetition (Gemini 1.5 Flash)]{\usebox\repetitionbox \label{code:repetition}
  }
  \hspace{0.04\linewidth}
  \subfloat[An example of ending generation with the inner end tag (Llama 3.2)]{\usebox\noclosetagbox \label{code:no-close-tag}} \\
  \subfloat[An example of ignoring the prompt (Mistral v0.3)]{\usebox\ignoringbox   \label{code:ignorant}}
  \caption{The actual output from the LLM/LMM with ``Extraction'' errors. We omitted parts of the output with three dots (\textbf{\dots}).}
  \label{fig:code-extraction-errors}
\end{figure*}
We show the output examples of the three ``Extraction'' error types in \Fref{fig:code-extraction-errors}. Note that SVG is a special type of XML, and SVG code should also comply with XML grammar. We explained the case of \Fref{fig:code-extraction-errors}\subref{code:repetition} in Sec 4.4. In this example, Gemini 1.5 Flash repeatedly generated \texttt{5 5,-5 5,}. Therefore, the generation ended before the model outputted the end tag (\texttt{</svg>}), making the code grammatically incorrect. This phenomenon was especially significant in open-sourced models. The code in \Fref{fig:code-extraction-errors}\subref{code:no-close-tag} ended with the close tag of the inner \texttt{<path>} tag, while it should finish with the end tag of the root element (\texttt{</svg>}). This code is also incorrect as XML and regarded as an ``Extraction'' error. \Fref{fig:code-extraction-errors}\subref{code:ignorant} shows an example of an LLM ignoring the prompt \texttt{Only return the SVG code}. Here, Mistral v0.3 explained the editing procedure and generated an SVG code snippet for each editing step. The output would be an ``Extraction'' error with our algorithm since this contained multiple code segments.

\subsection{Cases of ``Parsing'' errors}
\newsavebox\namespacebox
\newsavebox\attributebox
\newsavebox\commandbox
\newsavebox\argumentbox

\begin{lrbox}{\namespacebox}
  \begin{minipage}{0.96\linewidth}
    \begin{lstlisting}
<svg xmlns="http://www.w3.org/2000/svg" viewBox="0 0 36 36" style="background-color:purple; border-radius:50%;">
  <b#...#>
  <use <o#xlink:href="#reflector"#>/>
  <g>
    <b#...#>
  </g>
</svg>\end{lstlisting} 
\end{minipage}
\end{lrbox}

\begin{lrbox}{\attributebox}
  \begin{minipage}{0.96\linewidth}
    \begin{lstlisting}
<svg xmlns="http://www.w3.org/2000/svg" viewBox="0 0 36 36"><path <o#fill="#FFC5C5"#> d="M27.843 11.9c.205-.371.39-.766<b#...#>C29.519 20.484 30 18.943 30 17.462zM23.688 1.373c0-1.011-3.808-1.159-5.972-1.164H17.6c-2.164.005-5.972.152-5.972 1.164 0 2.013 3.796 2.942 5.887 2.982h.285c2.093-.041 5.888-.97 5.888-2.982z" <o#fill="#272B2B"#>/><b#...#></svg>\end{lstlisting} 
  \end{minipage}
\end{lrbox}

\begin{lrbox}{\commandbox}
  \begin{minipage}{0.96\linewidth}
    \begin{lstlisting}
<svg xmlns="http://www.w3.org/2000/svg" viewBox="0 0 36 36">
  <b#...#>
  <path fill="none" d="M35.885 <b#...#> 1.204-2.708 2.708z<o#F#>28.933 14.958a.064.064 0 0 1 .024.088.064.064 0 0 1 -.024-.088z">
</path>
</svg>
\end{lstlisting} 
  \end{minipage}
\end{lrbox}

\begin{lrbox}{\argumentbox}
  \begin{minipage}{0.96\linewidth}
    \begin{lstlisting}
<svg width="32" height="32" viewBox="0 0 32 32" fill="none" xmlns="http://www.w3.org/2000/svg">
<path <o#transform="matrix(-1 0.448,0 0.838,0 0.838-15)"#> d="M6 1C3.23079 1 1 3.23079 1 6V26C1 28.7614 3.23079 31 6 31H26C28.7614 31 31 28.7614 31 26V6<b#...#>" fill="#212121"/>
</svg>\end{lstlisting} 
  \end{minipage}
\end{lrbox}

\begin{figure*}
  \centering
  \subfloat[An example of output code using an undefined namespace. The output code is from Phi-3.5-mini.]{\usebox\namespacebox \label{code:undefined-namespace}
  }
  \hspace{0.04\linewidth}
  \subfloat[An example of the same attribute defined multiple times in the same element. The output code is from Llama 2.]{\usebox\attributebox \label{code:multiple-attributes}} \\
  \subfloat[An example of unknown line command included in the \texttt{d} attribute. The output code is from CodeGemma.]{\usebox\commandbox \label{code:unknown-command}}
  \hspace{0.04\linewidth}
  \subfloat[An example of the same attribute defined multiple times in the same element. The output code is from Llama 2.]{\usebox\argumentbox \label{code:too-many-arguments}}
  \caption{Examples of ``Parsing'' errors. We omitted parts of the output with three dots (\textbf{\dots}).}
  \label{fig:code-parsing-errors}
\end{figure*}
For the ``Parsing'' errors, we point out two patterns of failures in SVG rasterization or Chamfer distance calculation. \Fref{fig:code-parsing-errors}\subref{code:undefined-namespace} is an example of an XML error using an undefined namespace. In XML, namespaces (\texttt{xlink} in this case) should be defined by specifying the URI(\url{http://www.w3.org/1999/xlink}, \etc). However, the definition is not in the code in \Fref{fig:code-parsing-errors}\subref{code:undefined-namespace}. \Fref{fig:code-parsing-errors}\subref{code:multiple-attributes} is another XML error. Here, the code specifies the same attribute (\texttt{fill}) in the same \texttt{<path>} element. This is not allowed in XML and, therefore, regarded as an ``Parsing'' error.

The following two examples are errors in SVG. \Fref{fig:code-parsing-errors}\subref{code:unknown-command} is an example where an unknown line command exists within the \texttt{d} attribute of a \texttt{<path>} element. The SVG specification does not define the command \texttt{F} in the \texttt{d} attribute. Therefore, we could not recognize the accurate contour of the shape. The code in \Fref{fig:code-parsing-errors}\subref{code:too-many-arguments} contains an error in the \texttt{transform} attribute. The \texttt{matrix} command of the \texttt{transform} attribute allows the shape to rotate and scale simultaneously. The code specifies seven arguments, while the \texttt{matrix} command takes only six. Therefore, the pipeline could not rasterize the code, and we could not calculate the metrics.

\subsection{Examples of editing errors}
We could observe the following two characteristics of SVG editing with LLMs/LMMs. These editing failures are shown on the right side of \Fref{fig:examples}.
\begin{enumerate}
  \item The LLMs/LMMs tend to avoid modifying paths, even if the prompt asked them to change the object's shape. They may only rotate the shapes or add text directly to the image. The top-center instance in \Fref{fig:examples} is a sample. This task asked Gemini 1.0 Pro to change the ``7'' shape into a ``9'' (prompt: \texttt{Change the number "7" to "9" on the keycap.}). The model compromised by adding a ``9'' in text on top of the existing ``7'' shape.
  \item Processing \texttt{<path>} elements is very likely to fail. LLMs/LMMs only generate simple polygons or collapse the image by inappropriate parameter adjustments. For instance, in the top-right example in \Fref{fig:examples}, the task only requested LlaVa-NeXT to change the background color (prompt: \texttt{Change the background color to blue.}). Nevertheless, the LMM 	changed the ``UP!'' shape unnecessarily. This modification made the text unreadable. They also find it difficult to position those new elements aligning with the original image (\eg the bottom-center example in \Fref{fig:examples}).
\end{enumerate}

\section{ADDITIONAL RELATED WORK}
In this section, we present additional related works that we could not include in the main paper for the readers' reference. We also point to the papers presenting the LLMs/LMMs used in the experiment.

\subsection{Text-to-Vector generation}
\label{subsec:T2V}
Along with the advancements of methods to generate images representing the input text prompt in the raster domain (Text-to-Image; T2I), research on Text-to-Vector (T2V) has become increasingly popular. The mainstream of T2V methods is to define a loss function in the raster domain and run an optimization loop with a differentiable rasterizer~\cite{diffvg}. CLIPDraw~\cite{CLIPDraw} aims to generate drawings from a text description. They realize this by comparing the input text and the image under editing with the CLIP~\cite{clip} encoder. 
VectorFusion~\cite{vectorfusion} generates vector images by integrating Stable Diffusion~\cite{stable-diffusion} and using SDS loss~\cite{dreamfusion} instead of CLIP similarity. Zhang \etal~\cite{T2V-NPR} first learns the embeddings of paths using a VAE and then generates an initial image by optimizing this path representation.

NIVeL~\cite{NIVeL} questions these methods involving path optimization. They point out that SDS loss cannot model the graphic-dependent representation structure. Hence, they propose using an MLP that takes a point in 2D space and outputs the probability that a shape includes the point. With this process, it is possible to obtain an interpretable SVG representation by generating shapes for each layer corresponding to different colors.

\subsection{Utilizing LLMs for Vector Graphics Processing}
\label{subsec:LLM-for-SVG}
Vector graphics, including SVG, are described in a text file. Therefore, we can directly input an image as code into a recently evolving L\textbf{L}Ms---not just L\textbf{M}Ms. Several prior research exist that make LLMs process vector graphics, especially SVG.

Prior work has attempted to solve various SVG processing tasks using LLMs. For vectorizing raster images~\cite{S2VG2, starvector}, researchers focus on the generated SVGs being unreadable and restricted to paths. Therefore, they aim to generate SVG code with language models. In Text-to-Vector generation~\cite{gpt4-experiments, SVG-LLM}, experiments show that LLMs understand the concept of shapes and colors and can perform visual tasks. Papers on SVG understanding~\cite{SVG-LLM, svg-visualizations} demonstrate that the LLM can perform low-level visual analytic tasks. They showed this by giving LLMs input images as SVG code and asking questions about them.

\subsection{LLMs and LMMs used in the experiment}
The following is the list of LLMs and LMMs evaluated in Sec. 4 and its references.
\begin{itemize}
  \item \textbf{Open-sourced General-purpose LLMs}: Gemma 1.1~\cite{gemma-kaggle}, Gemma 2~\cite{gemma2-paper, gemma-kaggle}, Llama 2~\cite{llama2}, Llama 3, Llama 3.1, Llama 3.2~\cite{llama3}, Mistral v0.2, Mistral v0.3~\cite{mistral}, Phi-3.5-mini~\cite{phi3}
  \item \textbf{Open-sourced Code-specific LLMs}: CodeGemma~\cite{codegemma}, Code Llama~\cite{codellama}
  \item \textbf{LMMs}: LlaVA-NeXT~\cite{llava-next}, Phi-3.5-vision~\cite{phi3}, Qwen2-VL~\cite{qwen2-vl}
  \item \textbf{Close-sourced Models}: Gemini 1.0 Pro~\cite{gemini}, Gemini 1.5 Flash, Gemini 1.5 Pro~\cite{gemini-1.5}, GPT-3.5, GPT-4o, GPT-4o mini, o1-mini
\end{itemize}

\bibliographystyle{IEEEbib}
\bibliography{supp, main}